\documentclass[preprint,amsmath,amssymb,aps]{revtex4-2}
\usepackage{graphicx}
\usepackage{dcolumn}
\usepackage{bm}
\usepackage{xcolor}

\newcommand\zfig[1]{{\color{blue}#1}}

\begin{document}

\preprint{}

\title{On the Roughness of Two-Dimensional Materials}

\author{Pengjie Shi, Zhiping Xu}
\email{Corresponding author: xuzp@tsinghua.edu.cn (Z.X.)}

\affiliation{Applied Mechanics Laboratory, Department of Engineering Mechanics, Tsinghua University, Beijing 100084, China}

\date{\today}

\begin{abstract}
This study examines the roughness of mechanically cleaved edges in 2D crystals and glasses using molecular dynamics simulations with chemically accurate machine-learning force fields.
Our results show that ultra-flat armchair and zigzag edges can be achieved in graphene by aligning the loading direction with specific lattice orientations.
Deviations from these orientations create kinks between the atomically smooth armchair and zigzag segments, with increased irregularities when dynamic effects are considered.
Fracture mechanics analysis highlights the kinetic and dynamic factors contributing to crack deflection and edge roughening.
In three-atom-thick 2D silica crystals, the relationship between edge morphologies and cleavage conditions is modified by their bilayer structure and sublattice asymmetry.
In 2D silica glasses, this correlation is further disrupted by topological disorder.
These insights are crucial for minimizing edge roughness in 2D materials, which is essential for their performance in mechanical and electronic applications.

\end{abstract}
\maketitle
\newpage

\section*{Introduction}
The roughness of a surface or an edge is an important measure of its quality.
Pursuing the lowest roughness is one of the eternal interests in materials science and engineering, which finds its significance in a wide spectrum of applications from tribology to microelectronics.
Atomistically smooth surfaces and edges offer structural superlubricity by harnessing their incommensurability~\cite{hod2018structural}, templates for epitaxial growth of high-quality semiconductors~\cite{zhao2024ultrahigh}, and electrical contacts approaching the quantum limit~\cite{wang2011graphene,li2023approaching}.

The roughness of materials is shaped by their growth processes and fabrication techniques.
The physics behind the roughening or smoothing processes can be understood from thermodynamic or kinetic arguments.
From the thermodynamic perspective, surfaces or edges with the minimum energy densities are preferred during its growth process, provided the atoms can diffuse on them and reside on to the most energy-favorable sites.
However, the surface tension of finite-sized crystals typically induces faceting and the formation of polyhedrons governed by Wulff's construction principles, resulting in steps and kinks that are thermodynamically or kinetically controlled~\cite{pimpinelli1999physics}.
In post-growth material processing, mechanical polishing and physical or (electro)chemical etching are widely used to create smooth surfaces and edges down to the \AA ngstrom level~\cite{zhang2023general}.

Recent findings indicate that even minor edge defects in 2D crystals can significantly degrade material performance, thereby limiting their promising applications in elastic strain engineering (ESE)\,\cite{cao2020elastic,han2020large}.
Consequently, kinetic control of edge roughness has garnered considerable attention.
Following principles of fracture mechanics, controlled cleavage produces atomically smooth edges in 2D crystals~\cite{feng2023controlling}.
However, the underlying physics of achieving minimum roughness remains to be clarified.
It has been argued in the fracture mechanics community that a substance is considered truly brittle when its fracture toughness is twice the specific surface free energy of the solid, that is, $G_{\rm c} = 2\gamma$.
A ductile crystal fails by plastic shear, through the motion of dislocations, for example, and the measured values of $G_{\rm c}$ can rise to as much as $10^{5}\gamma$~\cite{kelly1967ductile}.
However, the value of $G_{\rm c}$ can be significantly larger than $2\gamma$ for the non-equilibrium nature of fracture, even for extremely brittle crystals that fail to be bond-breaking where a non-equilibrium measure $\Gamma$ is needed~\cite{lawn1993fracture,shi2024non}.
Kinetic control plays a crucial role.

Notably, for crystalline surfaces or edges with well-defined orientations, the atomic flatness at a length scale much larger than the lattice constants is not assured.
For a 2D crystal in the honeycomb lattice, a `chiral' edge is composed by armchair and zigzag segments \zfig{(Fig. 1a)}.
The edge roughness is defined by the numbers of consecutive segments $n_{\rm A}$ and $n_{\rm Z}$ and can be significant if they are large, while the orientation depends only on the ratio $n_{\rm A}/n_{\rm Z}$ \zfig{(Fig. 1b)}.
In contrast, in glassy materials with a random bonding network, the minimum roughness can be lower than the length of a single bond \zfig{(Fig. 1c)}.
However, the formation of local defects, such as voids, can increase the roughness \zfig{(Fig. 1d)}.
For these reasons, it is of practical interest to explore the lower bounds of or the minimum roughness in crystalline and glassy materials, which can guide the preparation of atomistically smooth surfaces with certain orientations.

In this study, we performed full-atom molecular dynamics (MD) simulations to investigate the edge roughening mechanisms in 2D materials, motivated by the simplicity of atomic-level structures and the recent surge in interest.
By resolving the evolution of atomic-level structures during the cleavage process at the first-principles level, we elucidated the thermodynamic and kinetic factors that govern the morphology and roughness of the resulting edges.
This research provides valuable insights into the fracture mechanics of 2D structures and advances material fabrication techniques aimed at achieving extreme or specified flatness.

\section*{Results}
Under stretch, shear, or tearing, graphene edges can be mechanically cleaved along specific macroscopic directions.
Considering the cleavage process as a problem of fracture, the edge roughness is controlled by the local kinetics of bond distortion and breakage, which includes crack deflection by mode mixing or microstructural heterogeneity, and maybe crack branching triggered by dynamical stability.

To explore the roughening process of 2D crystals at the atomic scale, MD simulations are performed using neural-network force field for fracture (NN-F$^{3}$) trained by density functional theory (DFT) calculations, which allows the use of large-scale models that are required for fracture mechanics studies.
2D models of $50$ nm $\times$ $15$ nm and $100$ nm $\times$ $30$ nm are constructed for graphene and silica, respectively \zfig{(Fig. 2a-d)}.
The model sizes are comparable in the number of unit cells since the lattice constants of 2D silica are approximately twice that of graphene. 
Edge pre-cracks are created perpendicular to the loading directions by assuming the mode-I condition.
For crystalline graphene, we find that the edge morphology evolves during propagation, which is characterized by the sequence ($n_{{\rm A}i}$,$n_{{\rm Z}i}$).
The stabilization of the ($n_{{\rm A}i}$,$n_{{\rm Z}i}$) patterns indicates that the size or boundary effects can be precluded in the discussion, which allows us to assess the kinetic control of edge roughness \zfig{(Fig. 2e,f)}.

The simulation results show that, under uniaxial tension along the close-to-armchair direction ($\theta_\mathrm{Z}<9.82^\circ$), only zigzag edges are cleaved.
Conversely, for tension along the close-to-zigzag directions ($\theta_\mathrm{Z}>25.87^\circ$), only armchair edges cleave.
The cleaved edges are thus atomistically smooth with a roughness of $0$.
Beyond these two boundaries, cracks deflect from the zigzag or armchair edges via nucleating lattice kinks due to the rise of a mode-II component of the stress field that twists the crack tip \zfig{(Fig. 2f)}.
The strength of this twist is quantified by a mode-mixing factor, $M = K_{\rm II}/K_{\rm I}$, and the critical values $M_{\rm c}$ can be identified from the simulation results \zfig{(Fig. 2e)}.

As the elastic driving force $G$ exceeds the fracture resistance $G_{\rm c}$ of the lattice, cracks start to grow.
In this condition, the relation between the direction of crack edge, $\theta_{\rm Edge}$, and $\theta_\mathrm{Z}$ can be determined from the maximum energy release rate (MERR) criterion as\,\cite{shi2024non}

\begin{equation}
\theta_{\rm Edge} = \arg\min\frac{G_{\rm c}\left(\theta_{\rm Edge}\right)}{\cos^2\left(\theta_\mathrm{Z}-\theta_{\rm Edge}\right)},
\end{equation}

\noindent where $G_{\rm c}\left(\theta_{\rm Edge}\right)=2\Gamma_{\rm A}\left[\sin\left(\theta_{\rm Edge}\right)+A_G\sin\left(30^\circ-\theta_{\rm Edge}\right)\right]$\ is the orientation-dependent fracture toughness.
The anisotropic factor of $A_G=\Gamma_{\rm Z}/\Gamma_{\rm A}=0.96$ is calculated by the non-equilibrium edge energy density $\Gamma$ as a practice measure of the fracture roughness ($G_{\rm c} = 2\Gamma$)\,\cite{shi2024non}. 
For $\theta_\mathrm{Z}\in[9.82^\circ,20.63^\circ]$, $n_{{\rm Z}i} = 1$, with $n_{{\rm A}i} \ge 1$.
For $\theta_\mathrm{Z}\in[20.63^\circ,25.87^\circ]$, $n_{{\rm A}i} = 1$, with $n_{{\rm Z}i}\ge 1$ \zfig{(Fig. 3a,c)}.
The predicted boundaries $\theta_{\rm Z}=9.9^\circ$ and $\theta_{\rm Z}=25.0^\circ$ by scanning the solutions of the MERR criterion (Eq. 1) agree well with the simulation results ($\theta_{\rm Z}=9.82^\circ$ and $\theta_{\rm Z}=25.87^\circ$).
Interestingly, the cracks proceed along the deflected direction only for a single step before the mode-mixing factor is effectively reduced, before $M_{\rm c}$.
These results with $n_{{\rm A}i} = 1$ or $n_{{\rm Z}i} = 1$ along the roughened edges imply a one-to-one correspondence between crack direction and edge roughness.
Specifically at $\theta_\mathrm{Z}=20.63^\circ$, we have both $n_{{\rm A}i} = 1$ and $n_{{\rm A}i} = 1$.
The crack advances along the direction of $\theta_{\rm Edge} = 19.11^\circ$ in this condition, resulting in a local minimum of the edge roughness in the range of $\theta_\mathrm{Z}\in[9.82^\circ,25.87^\circ]$.

MD simulations of 2D crystalline silica reveal trends in the sequence ($n_{{\rm A}i}$,$n_{{\rm Z}i}$) \zfig{(Figs. 2g,h and 3b,c)}.
However, the three-atom-thick structure results in more complex edge morphologies \zfig{(Fig. 1c)}.
At the tip of a propagating crack, the two Si-O bonds experiencing the greatest forces can be cleaved in both the upper and lower layers.
This asymmetry in bond cleavage and the resulting lattice distortion contribute to rougher edges compared to the atomistically sharp edges in graphene \zfig{(Fig. 3e-h)}.
The bilayer structure and the presence of dangling bonds in broken Si-O-Si units further exacerbate atomic-level structural irregularities.
Consequently, the limits $n_{{\rm A}i} = 1$ or $n_{{\rm Z}i} = 1$ are rarely achieved, leading to perturbed directional dependence of edge roughness with varying loading directions.
Quantitatively, the anisotropy in fracture toughness for 2D crystalline silica is more pronounced than in graphene, with $\Gamma_\mathrm{A}=1.004$ eV/\AA, $\Gamma_\mathrm{Z}=0.906$ eV/\AA, and $A_G=0.90$ from our NN-F$^3$ calculations, resulting in a shift of the low (high) roughness regions.
The predicted boundaries ($\theta_{\rm Z}=13.8^\circ$ and $28.1^\circ$) from the MERR criterion (Eq. 1) agree well with the simulation results \zfig{(Fig. 3c)}.

To turn on the inertial effects, we perform MD simulations with loading rates in the range of $10^{-3}$-$10^{-5}$ ps$^{-1}$ \zfig{(Fig. 4a-f)}.
The roughness of cleaved edges is modified in the dynamic fracture process.
An increase in the strain rate leads to a higher line density of kinks along the edges, while the overall direction of these edges remains almost unchanged.
The conclusion holds for 2D crystals of both graphene and silica, aligning with the common knowledge that the inertial effect induces a mixed-mode correction at the crack tip\,\cite{freund1998dynamic}.
However, our discussion on the dynamics will not be elaborated since an in-depth discussion needs much larger models where crack instabilities can be triggered and observed.

Compared to 2D crystals, the disordered bonding network in 2D glassy silica leads to amplified irregularities along cleaved edges.
The crack paths are shaped by lattice kinks and the disorder trapping effect, often guided by voids forming ahead of the crack tip \zfig{(Fig. 2i,j)}\,\cite{shi2024strength}.
These voids are nucleated due to stress concentration at the crack tip ($\sim r^{{-1/2}}$) and heterogeneity in the material strength due to the presence of lattice defects or topological ones (e.g., pentagons and heptagons).
Compared to the 2D silica crystals, crack deflection in the glassy structures is less sharp and often strongly distorted.
This irregularity in the edge profiles indicates a statistical nature of material strength.
The roughness of 2D glasses increases with the level of disorder, quantified by the standard deviation of ring distributions, $s = \sqrt{E\left[P^2\left(n\right)\right] - E^2\left[P\left(n\right)\right]}$.
Here, $n$ is the number of atoms in the minimal ring, and $E$ is the mean of its distribution $P\left(n\right)$.
For 2D silica, a transition from the crystalline to glassy states occurs at $s \approx s_{\rm c} = 0.45$, beyond which the alignment between the loading direction and lattice orientations becomes irrelevant.
At low $s$, the crack velocity approaches a constant, indicating that the effect of lattice trapping is negligible.
However, for $s > s_{\rm c}$, cracks can be trapped by the disorder.
As disorder increases, the instantaneous crack speed exhibits a distinct distribution pattern. At higher disorder levels, the cumulative probability density function (CDF) of the crack velocity converges to an exponential distribution \zfig{(Fig. 4g,h)}, indicating a less correlated nature.

\section*{Discussion and Conclusion}

The edge roughness (\emph{Ra}, the arithmetic average of absolute height deviations measured from the mean line, \zfig{Fig. 3}) in the 2D crystals and glasses depends on the loading direction and the lattice orientation.
Understanding these dependencies allows for better control of roughness during mechanical cleavage.
Firstly, selecting the direction of uniaxial tension is crucial in determining the edge roughness of 2D crystals and glasses with low disorder. By aligning the loading direction with low-energy edge motifs (zigzag or armchair), one can achieve nearly atomically flat edges within a narrow range. Outside this range, low roughness edges can be formed by lattice kinks joining the armchair and zigzag segments, with roughness decreasing as kink density increases.
Secondly, dynamic fracture processes such as crack deflection and bifurcation should be avoided at high strain rates.
At low loading rates, minimizing inertial effects ensures conditions for achieving minimum edge roughness.
Elastic anisotropy can be leveraged to surpass the fracture resistance anisotropy inherent in the underlying crystalline structures.
For example, creating wrinkles in 2D crystals forces cracks to advance in the direction with the highest stress intensity factor (SIF), perpendicular to the wrinkling patterns~\cite{feng2023controlling}.
Consequently, edges with specific orientations can be produced.
Notably, the evidence of $n_{\rm A} = 1$ or $n_{\rm Z} = 1$ under quasi-static loading conditions links the macroscopic crack orientation to the atomic-level edge structures and their associated roughness.

Our discussion extends to existing experimental facts of edge morphologies and roughness reported for 2D materials.
In graphene, edge morphologies and roughness are closely tied to processing techniques such as peeling\,\cite{qu2022anisotropic}, chemical vapor deposition (CVD)\,\cite{tian2011direct}, and dry or gas-phase etching\,\cite{zhao2021nanoimaging,wang2010etching}. 
Without controlling the loading conditions, the peeling method yields a roughness of approximately $4.8$ nm.
In contrast, the dry etching method can achieve roughness as low as $1.5$ nm. 
Additionally, edge morphology can be further disrupted by chemical asymmetry, as evidenced by the experimental observation of branched fractured edges in hexagonal boron nitride (h-BN)\,\cite{shi2024planning}.
By controlling loading directions, loading rates, or elastic anisotropy, atomically smooth edges can be achieved.
Regarding edge generation experiments for 2D glass, no relevant literature has been found to the best of our knowledge.
However, our results suggest that a low degree of disorder is favored to suppress edge roughness.

In summary, we elucidate the origins of edge roughening during the mechanical cleavage of 2D crystals and glasses.
Chemically accurate neural network force fields enable quantitative studies of atomistic fracture processes.
The anisotropy in fracture toughness of 2D crystals and the atomic-scale heterogeneity in 2D glasses lead to localized deflection of crack tips. 
Minimizing edge roughness can be achieved by carefully selecting loading directions and mitigating dynamic effects, particularly relevant for tribological and microelectronics applications.
Although our findings apply to 3D solids in general, the presence of kinks along terraces of cleaved surfaces introduces additional complexity.
Notably, the simulation results also provide valuable insights into fracture mechanics by delivering direct data on crack deflection and bifurcation~\cite{lawn1993fracture,liu2021analytical,liu2023semi}.

\section*{Models and Methods}
\subsection*{Atomic-level structural models}
To study the mechanical cleavage of 2D crystals and glasses, we choose graphene and 2D silica (SiO$_{2}$) as our model systems, which share the same honeycomb lattice but different chemistry.
A precrack is created along the $x$ direction.
Displacement-controlled uniaxial tensile tests are then performed along the $y$ direction.
The lattice orientation is measured by an angle $\theta$ between the $x$ direction and the nearest zigzag motif (Fig. \ref{Fig_2}a).
The cleaved edges are composed of armchair and zigzag segments.
A sequence of $\left(n_{\rm A1}, n_{\rm Z1}, n_{\rm A2}, n_{\rm Z2}, \cdots\right)$ is used to describe the edges, where $n_{\rm A}$ and $n_{\rm Z}$ are the numbers of consecutive armchair and zigzag structural units, respectively.

The atomic-level structures of 2D silica glasses are constructed by adapting a dual-switch procedure~\cite{ebrahem2020vitreous}, where the topological disorder is generated by $10^4$ Monte Carlo steps with a fictitious temperature of $10^{-4}$~\cite{font2022predicting}.
The product structures follow the Aboav-Weaire law with a model parameter of $\alpha = 1/3$ and agree with the experimental evidence, where small rings tend to sit around larger ones~\cite{ebrahem2020vitreous}.

\subsection*{MD simulations}

To capture the non-equilibrium nature of fracture processes, we use the chemically-accurate neural-network force field for fracture (NN-F$^{3}$) in our molecular dynamics (MD) simulations~\cite{shi2024planning,shi2024non,shi2024strength}.
In the workflow to construct NN-F$^{3}$ models, we pre-sample the 3D space of basal-plane strain states and introduce an active-learning workflow to explore the bond-breakage and (re)formation events during cracking using the end-to-end, symmetry-preserving DeepPot-SE framework\,\cite{NEURIPS2018_e2ad76f2} and DeePMD-kit\,\cite{deepmd}.
The pre-sampling technique boosts the efficiency in exploring the high-dimensional space of strain states, and the active-learning process ensures the accuracy of predictions by enriching the training dataset.
The technical details, scripts, and datasets can be found in our recent reports~\cite{shi2024planning,shi2024non,shi2024strength}.
The accuracy of NN-F$^{3}$ is confirmed to offer a root mean square error (RMSE) of $\le 2$ meV/atom for the energy per atom, $\le 50$ meV/\AA~for atomic forces, and $\le 30$ meV/\AA$^2$ for stress, respectively.

The generated NN-F$^{3}$ are sufficiently accurate for the study of mechanical cleavage.
For a typical fracture problem of 2D fracture, the local processes at the crack-tip and the far-field stress are concerned. 
As a result, the cubic scaling of first-principles (e.g., density functional theory or DFT) calculations hinders their applicability in simulating the fracture problem where a significant size effect exists.
In comparison, although the NN-F$^{3}$ simulations are $\sim 100$ times slower than the empirical force fields (FFs, e.g., Stillinger-Weber) for the models studied in this work, the computational cost is much lower than DFT calculations, and the same linear scaling to the number of atoms is maintained.

Quasi-static MD simulations are conducted to extract the stress-strain relation and fracture patterns.
The inertial effects are explored by MD simulations at specific loading rates.
The calculation of stress intensity factors (SIFs) follows Refs.\,\cite{over-determined,wilson_2019}, where the displacement field in an annular region around the crack tip (see the inset of Fig.\,\ref{Fig_2}c) is fitted to the Williams power expansion \cite{williams1957}.

\begin{acknowledgements}
This study was supported by the National Natural Science Foundation of China through grants 12425201 and 52090032. The computation was performed on the Explorer 1000 cluster system of the Tsinghua National Laboratory for Information Science and Technology.
\end{acknowledgements}

\newpage

\section*{Figures and Figure Captions}
\begin{figure*}[htp]
{
\includegraphics[width=\textwidth]{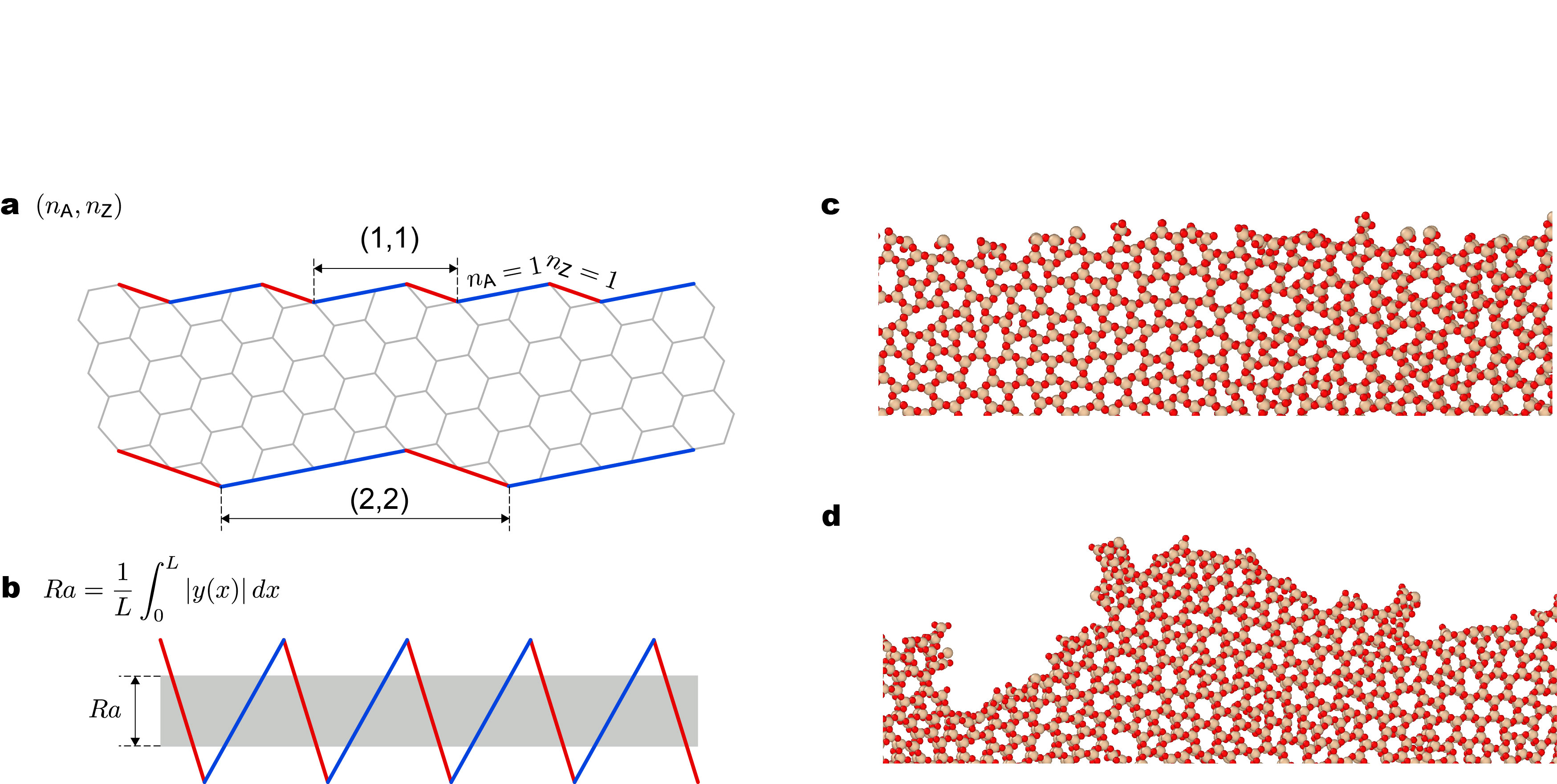}
\caption{
({\textbf a}) Crystalline edges of a honeycomb lattice, which are composed of armchair (blue) and zigzag (red) segments.
It should be noted that the same edge orientation does not necessarily correspond to the same edge roughness defined by $Ra$, the arithmetic average of absolute height deviations measured from the mean line ({\bf b}).
({\bf c}) Typical edges of an amorphous lattice with `minimum' roughness, displaying an irregular ring distribution.
({\bf d}) Rough edges of 2D glasses with irregular patterns including those created by voids nucleated during fracture.
}
\label{Fig_1}
}
\end{figure*}

\begin{figure*}[htb]
{
\includegraphics[width=\textwidth]{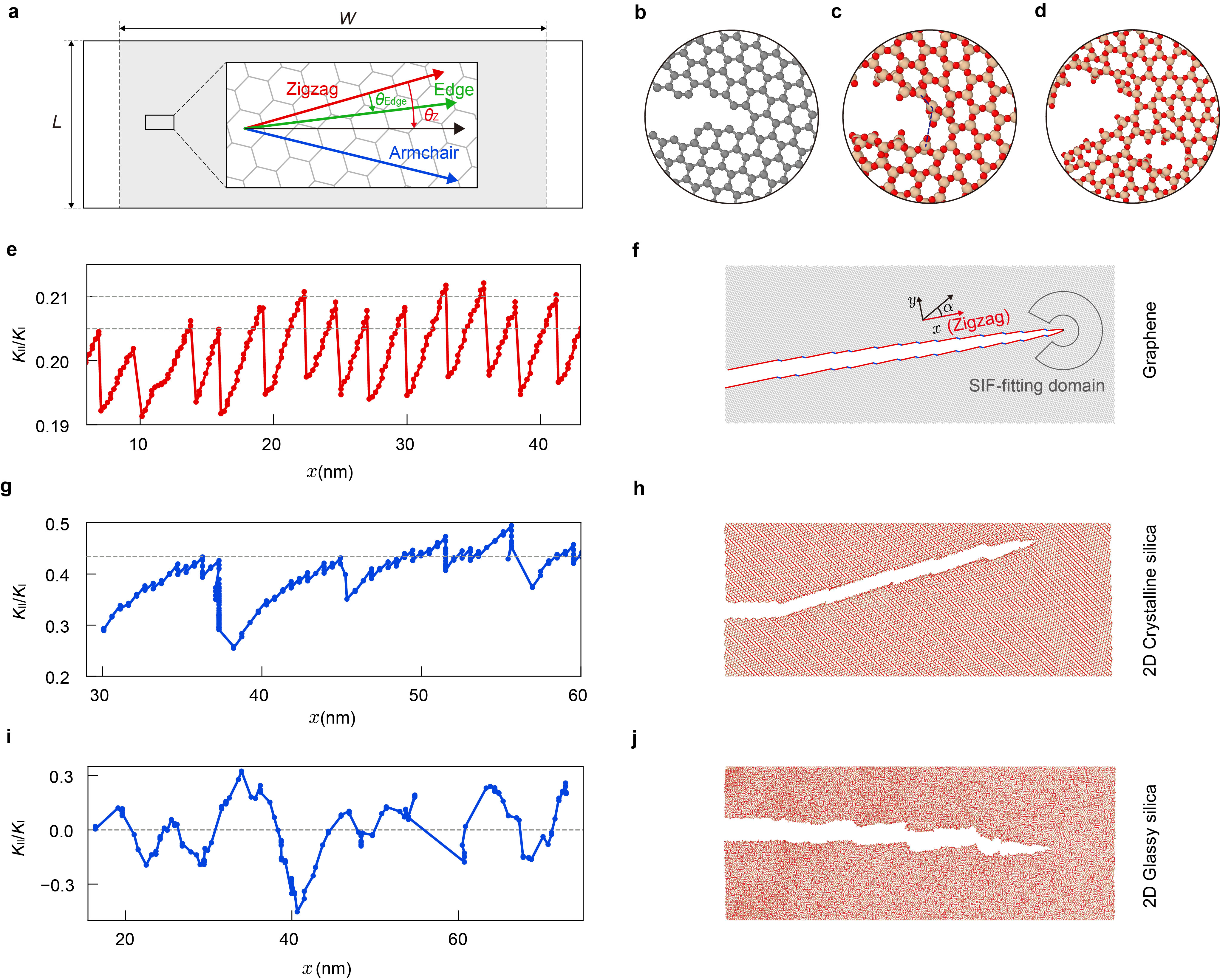}
\caption{
({\bf a}) Uniaxial tension setup in molecular dynamics (MD) simulations.
({\bf b-d}) Typical crack-tip configurations in graphene ({\bf b}), 2D silica crystals ({\bf c}) and glasses ({\bf d}).
({\bf e-j}) Mode-mixing factors ($K_{\rm II}/K_{\rm I}$, panels {\bf e}, {\bf g}, {\bf i}) and edge morphologies ({\bf f}, {\bf h}, {\bf j}) for graphene ({\bf e-f}), 2D silica crystals ({\bf g-h}) and glasses ({\bf i-j}).
The dash lines mark the range of critical mixing-mode factor $M_{\rm c}$, where the lower and upper bounds are specifically annotated for graphene.
}
\label{Fig_2}
}
\end{figure*}

\begin{figure*}[htp]
{
\includegraphics[width=\textwidth]{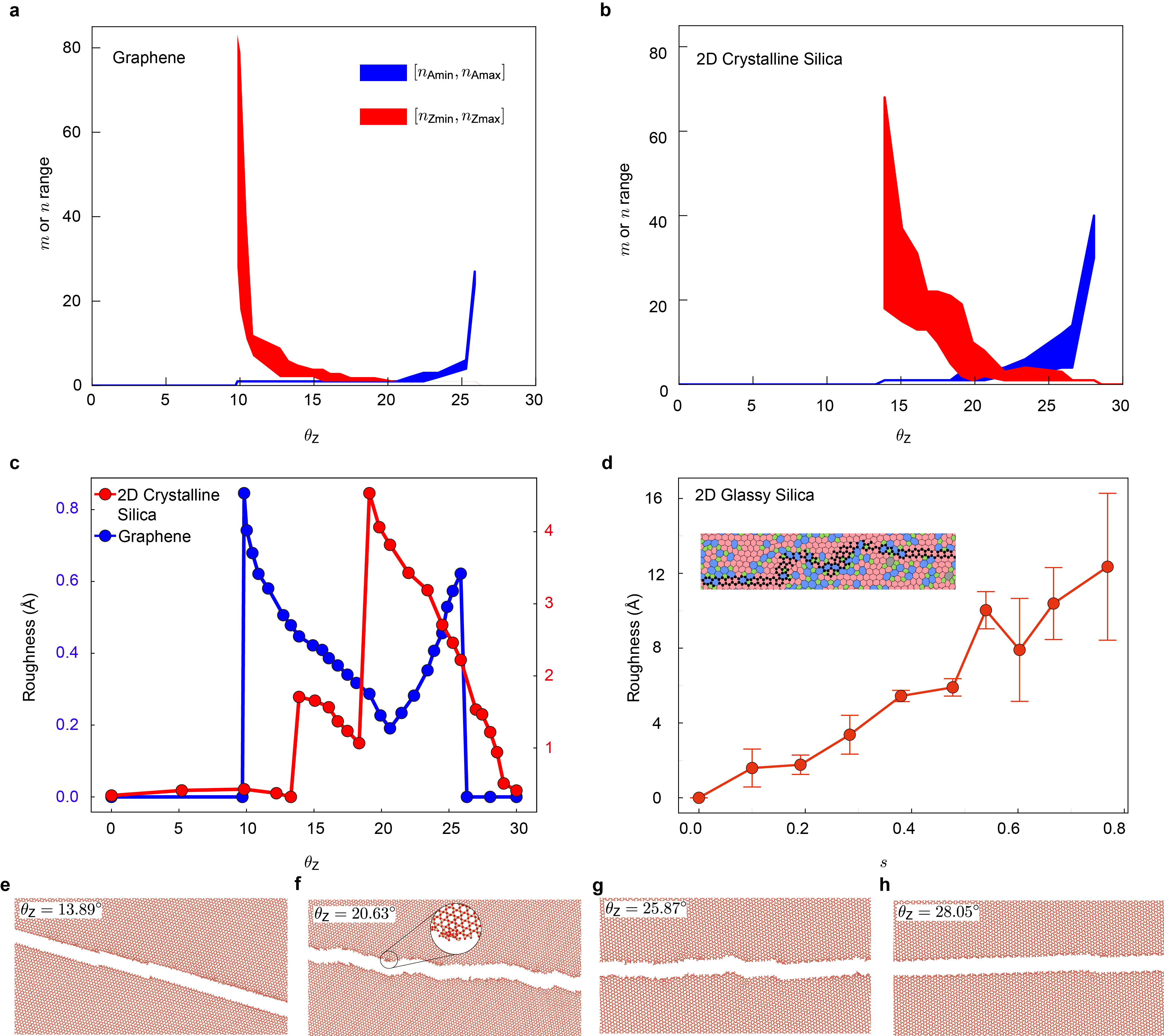}
\caption{
({\bf a-b}) Statistics of armchair and zigzag edge segments ($m$, $n$) the cleaved edges in graphene ({\bf a}) and 2D silica crystals ({\bf b}), where the samples are cleaved along direct orientation $\theta_{\rm Z}$.
({\bf c}) Dependence of edge roughness ($Ra$) on the loading direction for graphene and 2D silica crystals.
({\bf d}) Dependence of edge roughness on the degree of disorder of 2D silica glasses, defined as the standard deviation of the ring distribution,  $s = \sqrt{E\left[P^{2}\left(n\right)\right] - E^{2}\left[P\left(n\right)\right]}$.
({\bf e,f}) Edge morphologies of 2D silica crystals exhibiting irregularities due to their bilayer structure and sublattice asymmetry.
}
\label{Fig_3}
}
\end{figure*}

\begin{figure*}[htp]
{
\includegraphics[width=\textwidth]{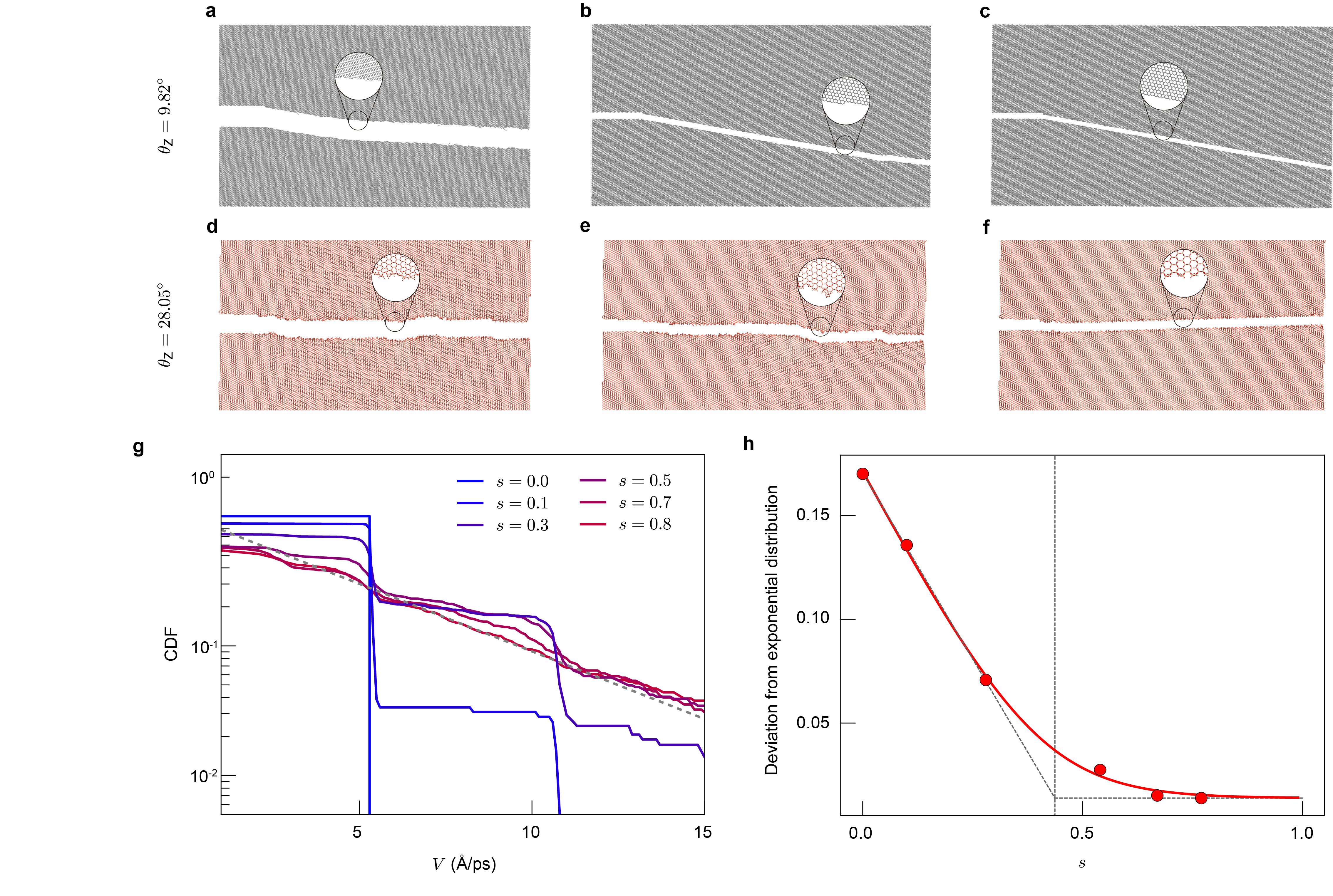}
\caption{
({\bf a-f}) Edge morphologies of graphene ({\bf a}-{\bf c}) and 2D silica crystals ({\bf d}-{\bf f}) cleaved under dynamic fracture.
The loading rates are $10^{-3}$ ({\bf a},{\bf d}), $10^{-4}$ ({\bf b},{\bf e}), and $10^{-5}$ ({\bf c},{\bf f}), respectively.
({\bf g,h}) Cumulative probability density function (CDF) of the instantaneous crack velocities at different levels of disorder, $s$ ({\bf g}), and the deviation from an exponential distribution, $d$ ({\bf h}).
}
\label{Fig_4}
}
\end{figure*}

\newpage
\clearpage

\bibliography{main_text}

\end{document}